\documentclass[12pt]{iopart}
\newcommand{\fant}[1]{\phantom{#1}}
\newcommand{\be}{\begin{equation}}
\newcommand{\ee}{\end{equation}}
\newcommand{\wdg}{\wedge}
\newcommand{\ot}{\otimes}
\begin{document}

\title[Minimal massive 3D gravity]{An alternative derivation of the minimal massive 3D gravity}
\author{Ahmet Baykal}
\address{Department of Physics, Faculty of Arts and Sciences, Ni\u gde University,  Bor Yolu,  51240 Ni\u gde, Turkey}
\ead{abaykal@nigde.edu.tr}
\begin{abstract}
By using the algebra of exterior forms and the first order formalism with constraints,
an alternative derivation of the field equations for the Minimal massive 3D gravity model is presented.
\end{abstract}

\maketitle

\section{Introduction and geometrical preliminary}

The Minimal massive gravity (MMG)  has recently  been introduced in \cite{MMG-I} as an alternative  to the Topologically massive gravity (TMG)  in three dimensions \cite{DJT}.  The field equations for MMG involve a particular quadratic-curvature terms obtained in a nontrivial way by extending the TMG Lagrangian.
The MMG model has the same gravitational degree of freedom as the TMG has and the linearization of the metric field equations for MMG yield a single propagating massive spin-2 field. At the same time, the complicated issue of matter coupling to the MMG model, in the particular form of an ideal fluid, is studied in \cite{MMG-II}. Compared to its cousin TMG, the new feature  of the MMG model is the positivity of the central charge defined for the  holographically dual conformal field theory  on a three dimensional anti-de Sitter ($AdS_3$) boundary. The hamiltonian analysis that follow from  the MMG Lagrangian  and the explicit computation of the dual conformal field theory  charge are also provided along with  the original derivation of the MMG equations \cite{MMG-I}. The present brief report deals with an alternative derivation of the MMG field equations using a tensorial language.

The organization of the paper is as follows. After introducing the geometrical notation in the rest of the introductory section, the MMG field equations are derived in the following section. The derivation of the field equations given here provides further insight into the complicated system of field equations that follow from MMG Lagrangian. The paper ends with a short concluding section commenting on matter coupling to the MMG model.
For convenience of the reader the basic geometrical definitions and quantities used below are summarized in the rest of the introductory section.

The geometrical notation for the exterior algebra required in the study of the MMG equations can be summarized as follows.
The metric tensor has constant components relative to an orthonormal coframe, $g=\eta_{ab}\, e^a\ot e^b$ with $\eta_{ab}=\mbox{diag}(-++)$ and the Latin indices refer to an orthonormal coframe that can be expanded into coordinate coframe as $e^a=e^{a}_{\mu}dx^\mu$. The set of basis frame fields is $\{X_a\}$ and the abbreviation $i_{X_a}\equiv i_a$ is used for the contraction operator with respect to the basis frame field $X_a$. $*$ denotes the Hodge dual operator acting on the basis forms, $*1=e^{0}\wdg e^1\wdg e^2=\frac{1}{3!}\epsilon_{abc}e^{abc}$ stands for the oriented volume element and $\epsilon_{abc}$ is the permutation symbol in three dimensions. The abbreviations, for example,  of the form $e^{ab}\equiv e^{a}\wdg e^b$ for the exterior products of basis 1-forms are  used for the convenience of the notation. The first structure equation  of Cartan reads
\be\label{cse1}
\Theta^a
=
D(\omega)e^a
=
de^a+\omega^{a}_{\fant{a}b}\wdg e^b
\ee
The first Bianchi identity can be written in the form $D(\omega)\Theta^a=D^2(\omega)e^a=\Omega^{a}_{\fant{a}b}(\omega)\wdg e^b$.
$D(\omega)$ is the covariant exterior derivative operator for the connection 1-form $\omega^{a}_{\fant{a}b}$, acting on tensor-valued forms, and a suitable definition and its relation to the covariant derivative $\nabla_a$ can be found in \cite{benn-tucker}.
The curvature 2-form $\Omega^{a}_{\fant{a}b}$ with
$\Omega^{a}_{\fant{a}b}(\omega)
=
\frac{1}{2}R^{a}_{\fant{a}bcd}(\omega)e^{cd}$ satisfies the Cartan's second structure equation
\be
\Omega^{a}_{\fant{a}b}(\omega)
=
d\omega^{a}_{\fant{a}b}
+
\omega^{a}_{\fant{a}c}\wdg \omega^{c}_{\fant{a}b}.\label{secondstr}
\ee
Ricci 1-form and the scalar curvature can be defined as the contractions $R^a=R^{ac}_{\fant{ac}bc}e^b\equiv R^{a}_{\fant{a}b}e^b=i_b\Omega^{ba}$ and $R=i_a R^a$, respectively. It is convenient to define 3D Einstein 2-form in terms of curvature 2-form as
\be\label{einstein-form-def}
*G^a(\omega)
=
-\frac{1}{2}\Omega_{bc}(\omega)*e^{abc}
=
-\frac{1}{2}\epsilon^{abc}\Omega_{bc}(\omega)
\ee
where $G^a(\omega)=R^a(\omega)-\frac{1}{2}R(\omega)e^a$ corresponding to  the connection 1-form $\omega^a_{\fant{a}b}$ in the same way as the
Einstein 2-form defined in the Riemannian context.  With a non-vanishing torsion, the first Bianchi identity takes the form $D(\omega)\Theta^a=\Omega^{a}_{\fant{a}b}(\omega)\wdg e^a$ and consequently Einstein tensor  is not  be symmetrical in Riemann-Cartan geometry
in general. The Levi-Civita connection is denoted by $\Gamma^{a}_{\fant{a}b}$ relative to an orthonormal coframe. With vanishing nonmetricity 1-form,
$Q_{ab}=-\frac{1}{2}D(\omega)\eta_{ab}=0$, the connection $\omega^{a}_{\fant{a}b}$  can be decomposed into the sum of a  pseudo-Riemannian  part and a contorsion
part as
\be
\omega^{a}_{\fant{a}b}
=
\Gamma^{a}_{\fant{a}b}+K^{a}_{\fant{a}b}
\ee
where the antisymmetric tensor-valued contorsion 1-forms $K_{ab}=-K_{ba}$ is related to the torsion 2-form by $\Theta^a=K^{a}_{\fant{a}b}\wdg e^b$. Consequently, the curvature 2-form $\Omega^{a}_{\fant{a}b}(\omega)$ can be decomposed as
\be\label{curv-decomp-general}
\Omega^{a}_{\fant{a}b}(\omega)
=
\Omega^{a}_{\fant{a}b}(\Gamma)+D(\Gamma)K^{a}_{\fant{a}b}+K^{a}_{\fant{a}c}\wdg K^{c}_{\fant{a}b}.
\ee
The decomposition of the curvature 2-form given in Eq.  (\ref{curv-decomp-general}) allows one to rewrite a given set of  field equations in the Riemann-Cartan geometry context in terms of  pseudo-Riemannian quantities \cite{hehl-heyde-kerlick-nester}.
It is well-known that in 3D there is no Weyl 2-form and the curvature 2-form of a Riemann tensor can be expressed in terms of its contractions
as
\be
\Omega^{ab}(\Gamma)
=
e^a\wdg L^{b}(\Gamma)-e^b\wdg L^{a}(\Gamma)\label{curv-id1}
\ee
where the Schouten 1-form $L_{a}=L_{ab}e^b$ can be defined in terms of Ricci 1-form and the scalar curvature as
\be
L^a
=
R^a-\frac{1}{4}Re^a,
\ee
($\Gamma$-dependence is omitted from the equation for convenience).
The Schouten 1-form can be used to derive the Cotton 2-form  $C^a=\frac{1}{2}C^{a}_{\fant{a}bc}e^{bc}$, explicitly one has $C^a=D(\Gamma)L^a(\Gamma)$.

The second Bianchi identity reads $D(\omega)\Omega^{a}_{\fant{a}b}(\omega)=0$ and for the curvature 2-form corresponding to the connection 1-form $\Gamma^a_{\fant{a}b}$, the identity $D(\Gamma)\Omega^{a}_{\fant{a}b}(\Gamma)=0$ yields  $e^a\wdg C^b=e^b\wdg C^a$ by using Eq. (\ref{curv-id1}).
The Cotton 2-form $C^a=\frac{1}{2}C^{a}_{\fant{a}bc}e^{bc}$ can be related to the symmetric traceless Cotton tensor $C_{ab}=C_{ba}$ by the relation
\be\label{cotton-tensor-form-rel}
C_{ab}
=
i_a*C_{b}.
\ee
By using the definition $C^a=D(\Gamma)L^a(\Gamma)$ and the relation $i_aDL_{bc}= \nabla_{a}L_{bc}$ in Eq. (\ref{cotton-tensor-form-rel}), one can obtain a more familiar expression for the Cotton tensor relative to an orthonormal coframe as
\be
C^{cd}
=
\epsilon^{abc}\nabla_{a}L^{d}_{\fant{a}b}
\ee
where $\nabla_a$ is covariant derivative corresponding to  the Riemannian connection $\Gamma^{b}_{\fant{a}c}$.

The properties of the Cotton 2-form has been previously studied using  the language of the tensor-valued exterior forms
and for further properties of the Cotton tensor and Cotton 2-form, for TMG and the other related gravitational models, the reader is referred to  \cite{cotton}.

\section{Minimal massive gravity Lagrangian and the field equations}

The field equations for the MMG model are derived from a Lagrangian 3-form
which can be  defined by a seemingly simple extension of the Lagrangian for TMG model with a cosmological constant.
The MMG Lagrangian, that is introduced recently in \cite{MMG-I}, can explicitly be rewritten in terms of exterior forms as
\be\label{MMG-lag}
L_{MMG}
=
L_{TMG}
+
\frac{\alpha}{2}\lambda_a \wdg \lambda_b\wdg *e^{ab}
\ee
where $\alpha$ is a coupling constant and the Lagrangian 3-form for the TMG Lagrangian \cite{DJT}  with a cosmological constant $\Lambda$ can be written in the form
\be
L_{TMG}
=
-
\frac{\sigma}{2}\Omega_{ab}\wdg *e^{ab}
+
\frac{1}{4\mu}(\omega^{a}_{\fant{a}b}\wdg d \omega^{b}_{\fant{b}a}
+
\frac{2}{3}\omega^{a}_{\fant{a}b}\wdg\omega^{b}_{\fant{b}c}\wdg \omega^{c}_{\fant{c}a})
+
\Lambda*1
+
\lambda_a\wdg \Theta^a.
\ee
In the TMG Lagrangian, the Einstein-Hilbert term is extended by the gravitational  Chern-Simons term with  constant $\mu$ and the auxiliary variable $\lambda^a=\lambda^a_{\fant{a}b}e^b$ is a vector-valued 1-form introduced to impose the vanishing torsion constraint for the TMG model. $\sigma$ is another
constant. For the motivation for introducing the $\lambda^2$-coupling term in the MMG Lagrangian, the reader is referred to the original reference \cite{MMG-I}.

As it stands, the MMG Lagrangian depends on three gravitational variables,
\be
L_{MMG}=L_{MMG}[e^a, \omega^{a}_{\fant{a}b},\lambda^a]\ee
 and
the   field equations for the variables $\{e^a\}, \{\omega^{a}_{\fant{a}b}\}, \{\lambda^a\}$ can derived from a variational principle
by using a first order formalism \cite{cotton,baykal-delice}. In contrast to the original derivation using the dualized connection 1-form, the present derivation makes use of the usual connection 1-form $\omega^{a}_{\fant{a}b}$ for the independent gravitational variable whose properties are defined in the preliminary section above.

One can show after some straightforward calculations in the exterior algebra that the total variational derivative of the Lagrangian (\ref{MMG-lag}) with respect to the independent variables is given by
\begin{eqnarray}
\hspace{-1em}\delta L_{MMG}
=&
\delta e_a\wdg
\left(
\sigma *G^a(\omega)
+
\Lambda*e^a+D(\omega)\lambda^a
+
\frac{\alpha}{2}\epsilon^{a}_{\fant{a}bc}\lambda^b\wdg \lambda^c
\right)
\nonumber\\
&+
\delta\omega_{ab}\wdg
\left\{
-
\frac{\sigma}{2}D(\omega)*e^{ab}
+
\frac{1}{2\mu}\Omega^{ba}(\omega)
-
\frac{1}{2}(e^a\wdg\lambda^b-e^b\wdg\lambda^a)
\right\}
\nonumber\\
&+
\delta\lambda_a\wdg
\left(
\Theta^a+\alpha\epsilon^{a}_{\fant{a}bc}e^b\wdg \lambda^c
\right)\label{total-var}
\end{eqnarray}
up to a disregarded boundary term. Here $\delta$ denotes the variation  of a quantity. For convenience, the technical details of the variational calculations leading to the important result (\ref{total-var}) are given in the appendix.

The terms on the right-hand side of the first line in (\ref{total-var}), namely, the variational derivative with respect to the basis coframe  1-forms give the metric field equations for the MMG Lagrangian whereas the second the third lines yield the equations of motion for the gravitational variable $\omega_{ab}$ and the auxiliary variable $\lambda_a$ respectively.

It is convenient to start with the field equations for the auxiliary field 1-form $\lambda^a$. Instead of imposing vanishing torsion  condition on the connection 1-forms $\omega_{ab}$, the equations  $\delta L_{MMG}/\delta\lambda_a\equiv P^a=0$ yields a torsion  in terms $\lambda^a$. Explicitly, by making use of the Cartan's first structure equations Eq. (1), the field equation
\be\label{torsion-mmg}
\Theta^a+\alpha\epsilon^{a}_{\fant{a}bc}e^b\wdg \lambda^c=0
\ee
can be rewritten in a more suggestive  form as
\be\label{conn-decomposition}
de^a+(\omega^{a}_{\fant{a}b}-\alpha\epsilon^{a}_{\fant{a}bc}\lambda^c)\wdg e^b=0.
\ee
Consequently,  the expression in the brackets in Eq. (\ref{conn-decomposition}) satisfy the structure equations with vanishing torsion.
Eq. (\ref{conn-decomposition})  for the Lagrange multiplier 1-form help to decompose the connection 1-form $\omega^{a}_{\fant{a}b}$ into
the Riemannian part denoted by $\Gamma^{a}_{\fant{a}b}$ defined by the second term  in (\ref{conn-decomposition}) and a contorsion part as
\be\label{conn-decomp-explicit}
\omega^{a}_{\fant{a}b}
=
\Gamma^{a}_{\fant{a}b}+\alpha\epsilon^{a}_{\fant{a}bc}\lambda^c
\ee
and one can readily identify the contorsion 1-form as $K^{a}_{\fant{a}b}=\alpha\epsilon^{a}_{\fant{a}bc}\lambda^c$.
The decomposition of the connection 1-form (\ref{conn-decomp-explicit}) can be used to decompose the curvature 2-form as well. Using the identity given in Eq. (\ref{curv-decomp-general}), one finds
\be
\Omega^{a}_{\fant{a}b}(\omega)
=
\Omega^{a}_{\fant{a}b}(\Gamma)
+
\alpha\epsilon^{a}_{\fant{a}bc}D(\Gamma)\lambda^c
-
\alpha^2\lambda^a\wdg\lambda_b.\label{curv-decomp-explicit}
\ee
Likewise, the Einstein 2-forms also decompose as
\be\label{eisntein-form-decomp-explicit}
*G^a(\omega)
=
*G^a(\Gamma)
-
\alpha D(\Gamma)\lambda^a
+
\frac{\alpha^2}{2}\epsilon^{a}_{\fant{a}bc}\lambda^b\wdg \lambda^c.
\ee

With the help of Eqs. (\ref{curv-decomp-explicit}) and (\ref{eisntein-form-decomp-explicit}), the connection equations can be expressed in terms of the Levi-Civita connection $\Gamma^{a}_{\fant{a}b}$. To this end, note first that by using the expression for the contorsion 1-form, one finds that the covariant exterior term can be written in terms of the auxiliary field 1-form as
\be
D(\omega)*e^{ab}
=
\epsilon^{ab}_{\fant{ab}c} \Theta^c
=
\alpha(e^a\wdg \lambda^b-e^b\wdg \lambda^a).
\ee
Thus, by combining these results, the vacuum field equations for the connection 1-forms obtained by $\delta L_{MMG}/\delta\omega_{ab}\equiv S^{ab}=0$  can be rewritten in the form
\be\label{reduced-conn-eqns}
-
\frac{1}{2\mu}\Omega^{ab}(\Gamma)
-
\frac{\alpha}{2\mu}\epsilon^{ab}_{\fant{ab}c}D(\Gamma)\lambda^c
+
\frac{\alpha^2}{2\mu}\lambda^a\wdg \lambda^b
-
\frac{1}{2}(1+\alpha\sigma)
(e^a\wdg\lambda^b-e^b\wdg\lambda^a)=0
\ee
in terms of the Riemannian quantities.

In contrast to the TMG case with $\alpha=0$, in which $\lambda^a$ can uniquely be solved, the connection equations (\ref{reduced-conn-eqns}) are not a set of algebraic equations in the MMG case. To put it more precisely, owing to the presence of the term
\be
D(\Gamma)\lambda^a
=
d\lambda^a+\Gamma^{a}_{\fant{a}b}\wdg \lambda^b
\ee
that contains the exterior derivatives of the auxiliary 1-form field $\lambda^a$,  the reduced equations  for the connection 1-forms yield  a set of dynamic equations for the auxiliary field $\lambda^a$. In the TMG case ($\alpha=0$) it is easy to find that the algebraic  equation (\ref{reduced-conn-eqns}) for $\lambda^a$ has a unique solution
\be\label{tmg-lag-multiplier}
\lambda^a
=
-\frac{1}{\mu}L^a(\Gamma)
\ee
with the help of the curvature identity (\ref{curv-id1}).
On the other hand, for $\alpha\neq0$, i.e., for the MMG case, it is difficult to find \emph{the} general solution for $\lambda^a$ in a closed form using the reduced connection equations given in (\ref{reduced-conn-eqns}). Consequently, it is not possible to eliminate the auxiliary vector-valued 1-form   $\lambda^a$
from the metric equations obtained by the coframe variations, $\delta L_{MMG}/\delta e^a\equiv *E^a=0$, in favor of the remaining gravitational variables. Thus, without a closed  expression for the Lagrange multiplier 1-form $\lambda^a$ in terms of other fields, it is not possible to reduce the MMG Lagrangian (\ref{MMG-lag}) to a form $L_{MMG-red.}[\Gamma^{a}_{\fant{a}b}, e^a]$ by a back-substitution.

With the benefit of the hindsight and the fact that for $\alpha=0$ the solution is of the form (\ref{tmg-lag-multiplier}), one can try a simple ``ansatz" proportional to the Schouten 1-form. However, assuming a non-vanishing cosmological constant, a suitable ansatz with two adjustable constant $p$ and $q$, is of a slightly more general  form:
\be\label{ansatz}
\lambda^a
\equiv
pL^a(\Gamma)+qe^a.
\ee
By plugging the ansatz (\ref{ansatz}) into Eq. (\ref{reduced-conn-eqns}),
by making use of the identity (\ref{curv-id1}), the definition of Cotton 2-form, $D(\Gamma)L^a(\Gamma)=C^a$,
and the torsion-free condition, $D(\Gamma)e^a=0$ for $\Gamma$, one finds
\be\label{reduced-conn-form2}
-\frac{1}{2}A\Omega^{ab}(\Gamma)
+
\frac{1}{2}B \epsilon^{ab}_{\fant{ab}c}C^c
+
\frac{1}{2}Ee^a\wdg e^b
+
\frac{1}{2}F L^a(\Gamma)\wdg L^b(\Gamma) =0
\ee
with the constants $A, B, E, F$ that can be expressed in terms of the constants  $p,q$  and the constants in the MMG Lagrangian  as
\begin{eqnarray}
&A
=
\frac{1}{\mu}
(\alpha^2pq-1)
-
p(1+\alpha\sigma),
\qquad
&B
=
-\frac{\alpha}{\mu},
\\
&E
=
\frac{\alpha^2q^2}{\mu}-2q(1+\alpha\sigma),
&F
=
\frac{p^2\alpha^2}{\mu}.
\end{eqnarray}

By multiplying  Eq. (\ref{reduced-conn-form2}) with a permutation symbol and taking the definition of the Einstein 2-form (\ref{einstein-form-def}) into account,   Eq. (\ref{reduced-conn-form2}) can finally be brought to the form, similar to the equation for basis  coframe 1-forms, as
\be\label{MMG-original form1}
A*G^a+BC^a+E*e^a+\frac{1}{2}F\epsilon^{a}_{\fant{a}bc}L^b\wdg L^c=0
\ee
with the obvious $\Gamma$-dependence of the curvature terms omitted.
It is possible to show that Eq. (\ref{MMG-original form1}) is equivalent to the expression for the MMG field equations  given in the original form relative to a coordinate basis \cite{MMG-I}. In particular, the quadratic-curvature terms can be expressed with a suitable vector-valued 2-form defined by
\be\label{J-def}
J^a
\equiv
\frac{1}{2}\epsilon^{a}_{\fant{a}bc}L^b\wdg L^c
\ee
in the notation used above. The original definition of the quadratic-curvature term corresponds to the coordinate components of the expression $J^{ab}\equiv i^a*J^b$. The first three terms in (\ref{MMG-original form1}) are the terms appearing in the TMG field equation (with cosmological constant) with shifted constants. The last term in (\ref{MMG-original form1}) is involves the quadratic-curvature terms, which is a consequence of the MMG field equations and as well as the ansatz (\ref{ansatz}). As it has been noted in the original construction \cite{MMG-I}, note that the trace of the quadratic-curvature term yields the quadratic-curvature part of the New massive gravity Lagrangian \cite{NMG} up to a constant multiple. By a straightforward calculation, one can show explicitly that
\be
e^a\wdg J_a
=
\frac{1}{2}L_a\wdg L_b\wdg *e^{ab}
=
-
\frac{1}{4}(R^a\wdg *R_a-\frac{3}{8}R^2*1)
\ee
where the curvature components in the expression correspond to the curvature of a Levi-Civita connection $\Gamma^{a}_{\fant{a}b}$.

Finally, the significance of the ansatz (\ref{ansatz}) becomes more pronounced after one rewrites the coframe (equivalently, the metric) equations
\be
\sigma *G^a(\omega)
+
\Lambda*e^a+D(\omega)\lambda^a
+
\frac{\alpha}{2}\epsilon^{a}_{\fant{a}bc}\lambda^b\wdg \lambda^c=0
\ee
in terms of the Riemannian quantities as well. By using the decomposition formulas (\ref{curv-decomp-explicit}) and (\ref{eisntein-form-decomp-explicit}), one finds that
\be\label{coframe-reduction1}
\sigma*G^a(\Gamma)
+
(1-\alpha\sigma)D(\Gamma)\lambda^a
+
\Lambda*e^a
+
\frac{1}{2}\alpha(3+\alpha\sigma)\epsilon^{a}_{\fant{a}bc}\lambda^b\wdg \lambda^c=0.
\ee

Now, assuming  that the coefficients  the terms  in Eq. (\ref{coframe-reduction1}) do not vanish
identically, one can now plug in the ansatz into Eq. (\ref{coframe-reduction1}).
Explicitly,  by plugging the ansatz (\ref{ansatz}) into the coframe equations,
and subsequently using the curvature identity (\ref{curv-id1}) and  the definition of the  quadratic-curvature term  $J^a$ given in Eq. (\ref{J-def}), one eventually finds that Eq. (\ref{coframe-reduction1}) can be rewritten in the form
\begin{eqnarray}
\left[\sigma-pq\alpha(3+\alpha\sigma)\right]*G^a
+
p(1-\alpha\sigma)C^a
+
&\left[\Lambda+q^2\alpha(3+\alpha\sigma)\right]*e^a
\nonumber\\
&
\fant{aaaa}+
p^2\alpha(3+\alpha\sigma)J^a=0\label{coframe-eqns-2}
\end{eqnarray}
with the $\Gamma$-dependence of the curvature terms omitted.

By comparison, one can  see that Eq. (\ref{coframe-eqns-2}) has the same form as that of  Eq. (\ref{MMG-original form1}) up to the constants multiplying each term. The ansatz  in Eq. (\ref{ansatz}) is unique in the sense that
it renders the coframe and the connection equations  identical up to a duality provided that the constants in the Eqs. (\ref{MMG-original form1}) and (\ref{coframe-eqns-2}) are identified accordingly.

Explicitly, in the present notation and relative to an orthonormal coframe, MMG equations
(using the constants with which they are defined originally) for the coframe and the connection 1-forms take the form
\be
\sigma*G^a+\Lambda_0*e^a+\frac{1}{\mu}C^a+\frac{\gamma}{\mu^2}J^a=0
\ee
that is an equation for 2-forms in a form in line with, for example, the form of the TMG field equations given in \cite{cotton} and  \cite{dereli-tucker-3D}.

Finally, the following remarks are in order regarding the derivation of the MMG field equations.
\begin{itemize}

\item[(i)] It is worth emphasizing that, without the assumption of the ansatz (\ref{ansatz}) for  $\lambda^a$, the field equations for the independent coframe  and the connection 1-forms  can be expressed in terms of pseudo-Riemannian quantities with the help of the constraint (\ref{torsion-mmg}). The connection equations (\ref{reduced-conn-eqns}) can be regarded as general set of dynamic equations for the auxiliary vector-valued 1-form field $\lambda^a$ and for the particular ansatz (\ref{ansatz}), the MMG field equations simplify considerably. To put it in mathematical terms, evaluated using the ansatz (\ref{ansatz}) in terms of a Levi-Civita connection,  the coframe equations  ${\delta L_{MMG}}/{\delta e_a}=*E^a$ can be made equal to  the dual of  the connection equations ${\delta L_{MMG}}/{\delta \omega_{ab}}=S^{ab}$ as
    \be
    \left.\frac{1}{2}\epsilon^{a}_{\fant{a}bc}S^{bc}\right|_{ansatz}=\left.*E^a\right|_{ansatz}
    \ee
    by adjusting the parameters of the MMG model and the ansatz.
\item[(ii)] Even though the MMG Lagrangian (\ref{MMG-lag}) contains a constraint term imposing the  vanishing torsion condition on the independent connection 1-form, it leads to a Riemann-Cartan type geometry \cite{hehl-heyde-kerlick-nester} and the MMG model is unique in the sense that, written in terms of Riemannian quantities, the equations for the connection and coframe  1-forms can be identified by a particular choice of the auxiliary field variable.
    This insight  for the field equations of the MMG model seems to be new. In $D>3$ dimensions,  the construction will not work for the reason that the derivation makes the essential use of the curvature identity (\ref{curv-id1}) peculiar to three dimensions.

\item[(iii)] In contrast to the original derivation of the MMG field equations that makes use of the dualized connection and the curvature forms, the relation given in Eq. (2.11)  in \cite{MMG-I}, namely,
$
e\cdot h=e^ah^{b}\eta_{ab}=0
$
is nowhere assumed  to hold in the derivation of the field equations obtained  from  Eq. (\ref{total-var}), but it is a simplifying property encoded in  the ``ansatz" given in Eq. (\ref{ansatz}).  As it is emphasized above, a general closed expression for $\lambda^a$ in terms of the other gravitational variables is not available, and hence it is  not possible to eliminate the auxiliary variable $\lambda^a$ from the metric equations or to reexpress the MMG Lagrangian  3-form in terms of the Riemannian quantities.

\item[(iv)]
Throughout  the above derivation, the tensorial manipulations  are preformed relative to an orthonormal coframe  yielding the field equations  relative to a orthonormal basis in a unified manner.
\end{itemize}

\section{Concluding comments}

With regard to the matter coupling of the MMG, it may be more convenient to consider the MMG field equations in terms of the
variables  $e^a, \omega^{a}_{\fant{a}b},\lambda^a$ before casting the field equations in a form expressed in terms of the
pseudo-Riemannian quantities.

It is possible to show that, in the first order formalism with independent connection and coframe forms,
the diffeomorphism invariance of a gravitational model leads to the more general differential identity \cite{kopczynsky}, involving the
covariant exterior derivative of the coframe equations.
Explicitly, for a general Lagrangian of the form $L=L[e^a,\omega^{a}_{\fant{a}b}, \Theta^a, \Omega^{a}_{\fant{a}b}]$
the invariance under a diffeomorphism $\phi$ can be expressed as
\be\label{diff-inv1}
(\phi^*L)[e^a,\omega^{a}_{\fant{a}b}, \Theta^a, \Omega^{a}_{\fant{a}b}]
=
L[\phi^*e^a,\phi^*\omega^{a}_{\fant{a}b}, \phi^*\Theta^a, \phi^*\Omega^{a}_{\fant{a}b}]
\ee
where $\phi^*$ stands for induced pullback map that  $\phi$ generates on the tensor-valued $p$-forms.
In particular, if one assumes that $\phi$ is generated by a vector field $Z$ depending on a parameter, say $t$, then taking the
derivative of (\ref{diff-inv1}) with respect to the parameter $t$, and using the definition of the Lie derivative $\mathcal{L}_Z$,
one readily finds
\be\label{noether-derivation2}
\mathcal{L}_ZL
=
(\mathcal{L}_Ze^a)\wdg *E^a+ (\mathcal{L}_Z\omega^{a}_{\fant{a}b})\wdg S^{ab}
\ee
where an exact 3-form  on the right-hand side is discarded.
Furthermore, by using the Cartan's formula, $\mathcal{L}_Z=di_Z+i_Zd$,  for  Lie derivative  with respect to the  vector field $Z=Z^aX_a$, and the Cartan's structure equations, one can simplify the expression in Eq. (\ref{noether-derivation2}) to the form
\begin{eqnarray}
\hspace{-3.5em}(\mathcal{L}_Ze^a)\wdg *E^a+ (\mathcal{L}_Z\omega^{a}_{\fant{a}b})\wdg S^{ab}
=&
-(i_Ze^a) D*E_a+i_Z\Theta_a\wdg *E^a
+
(i_Z\Omega_{ab})\wdg S^{ab}
\nonumber\\
&
+(i_Z\omega_{ab})\left\{\frac{1}{2}(e^a\wdg *E^b-e^b\wdg *E^a)-DS^{ab}\right\}
\nonumber\\
&
+d\left\{(i_Ze^a) *E_a+ (i_Z\omega_{ab})S^{ab}\right\}.\label{noether-derivation1}
\end{eqnarray}
Finally, assuming that the boundary term vanishes, and setting  the coefficients of $Z^a$ and $i_Z\omega_{ab}$ separately in (\ref{noether-derivation1}), one ends up with the identities
\begin{eqnarray}
&D*E_a
=
(i_a \Theta_{b})\wdg *E^b+(i_a \Omega_{bc})\wdg S^{bc},
\label{diff-invvariance-D-id}\\
&DS^{ab}
=
\frac{1}{2}(e^a\wdg *E^b-e^b\wdg *E^a),\label{local-lorentz-inv}
\end{eqnarray}
respectively.

In the above framework, the diffeomorphism invariance of a general Lagrangian generates the generalized differential identity given in Eq. (\ref{diff-invvariance-D-id}) that reduces $D*E^a=0$ for the  pseudo-Riemannian case whereas
the identity in Eq. (\ref{local-lorentz-inv}) resulting  from the invariance under coframe rotations (local Lorentz invariance)
reduces to the symmetry property $E_{ab}=E_{ba}$ expressed in the form $e^a\wdg *E^b-e^b\wdg *E^a=0$.

For the particular case of the MMG Lagrangian, the implementation of the auxiliary field $\lambda^a$  and the ansatz (\ref{ansatz}) into  the generalized
differential Bianchi identity  (\ref{diff-invvariance-D-id})  obtained by the Noether procedure requires further scrutiny. In this regard, the first order formalism  and Riemann-Cartan geometry may provide an alternative mathematical framework  in the construction of  consistent minimal matter coupling to the MMG model as well.

\ack
The author would like to thank   O. Teoman Turgut and \"Ozg\"ur Delice for instructive correspondences and help.

\section*{Appendix}

In the first order formalism, the independent gravitational variables $\{e^a\}$, $\{\omega^{a}_{\fant{a}b}\}$ and their first order derivatives $\{de^a\}$, $\{d\omega^{a}_{\fant{a}b}\}$ are allowed  in the Lagrangian 3-form. The derivatives of the variables appear only through tensorial quantities $\Theta^a$ and $\Omega^{a}_{\fant{a}b}$, respectively. Hence, it is convenient to assume that
$L_{MMG}=L_{MMG}[e^a, \omega^{a}_{\fant{a}b}, \Theta^a, \Omega^{a}_{\fant{a}b},\lambda^a]$ where the Lagrange multiplier vector-valued 1-form $\lambda^a$
is an auxiliary variable and impose vanishing torsion constraint for the connection in  the $L_{TMG}$ Lagrangian.

The field equations can be obtained by the principle of stationary  action, embodied in  the variational equation $\delta I_{MMG}=0$. For the MMG model, the action functional is given by the integral of the Lagrangian 3-form as
\be
I_{MMG}
=
\int_{U}L_{MMG}
=
\int_{U}L_{TMG}
+
\frac{\alpha}{2}\int_{U}\lambda_a \wdg \lambda_b\wdg *e^{ab}
\ee
where $U\subset M$ is an open subset on some chart defined on a (2+1)-dimensional Riemann-Cartan manifold $M$.

For convenience of the notation, the integral sign will be omitted and the variational derivative of the Lagrangian 3-form  will be considered. By using the product rule for the variational derivative,  $\delta L_{TMG}$ can explicitly be written as
\begin{eqnarray}
\hspace{-1em}\delta L_{TMG}
=&
-\frac{\sigma}{2} \delta\Omega_{ab}\wdg*e^{ab}
-
\frac{\sigma}{2}\Omega_{ab}\wdg\delta*e^{ab}
+
\delta\lambda_a\wdg \Theta^a+ \lambda_a\wdg \delta\Theta^a
+
\Lambda\delta*1
\nonumber\\
&
+
\frac{1}{4\mu}
\left(
\delta \omega_{ab}\wdg d\omega^{ba}
+
\omega_{ab}\wdg d\delta \omega^{ba}
+
2\delta \omega^{a}_{\fant{a}b}\wdg \omega^{b}_{\fant{b}c}\wdg \omega^{c}_{\fant{c}a}\label{var-expression1-appendix}
\right),
\end{eqnarray}
where the cyclic property
$
\delta \omega^{a}_{\fant{a}b}\wdg \omega^{b}_{\fant{b}c}\wdg \omega^{c}_{\fant{c}a}
=
\omega^{c}_{\fant{c}a}\wdg\delta \omega^{a}_{\fant{a}b}\wdg \omega^{b}_{\fant{b}c}
=
\omega^{b}_{\fant{b}c}\wdg\omega^{c}_{\fant{c}a}\wdg\delta \omega^{a}_{\fant{a}b}
$
is used to derive the expression in the second line.
To evaluate the variational derivative in Eq. (\ref{var-expression1-appendix}) further,  it is convenient to  recall the following variational derivatives  for various tensor-valued forms:
\begin{eqnarray}
&\delta *e^{ab}
=
\epsilon^{ab}_{\fant{ab}c}\delta e^c
=
\delta e^c*e^{ab}_{\fant{ab}c}
\label{var-id-1}\\
&\delta*1
=
\frac{1}{6}\epsilon_{abc}\left(\delta e^{a}\wdg e^b\wdg e^c+e^{a}\wdg\delta e^b\wdg e^c+ e^{a}\wdg e^b\wdg\delta e^c\right)
=
\delta e^a\wdg *e_a
\\
&\delta \Theta^a
=
\delta de^a+\delta\omega^{a}_{\fant{a}b}\wdg e^b+\omega^{a}_{\fant{a}b}\wdg \delta e^b
=
D\delta e^a - \delta e^b \wdg \omega^{a}_{\fant{a}b}
\\
&\delta\Omega^{a}_{\fant{a}b}
=
\delta d\omega^{a}_{\fant{a}b}
+
\delta\omega^{a}_{\fant{a}c}\wdg \omega^{c}_{\fant{a}b}+\omega^{a}_{\fant{a}c}\wdg \delta\omega^{c}_{\fant{a}b}
=
D\delta\omega^{a}_{\fant{a}b}.\label{var-id-4}
\end{eqnarray}
Note that the variational derivative $\delta$ commutes with the exterior derivative $d$,  however, it does not commute with the covariant exterior derivative $D$ and the Hodge dual. $\delta\omega^{ab}$ is a tensor and thus the covariant exterior derivative $D\delta\omega^{ab}$ can be defined as in the  expression on the right-hand side of Eq. (\ref{var-id-4}).

By using the variational formulas  given in Eqs. (\ref{var-id-1})-(\ref{var-id-4}) and the Cartan's structure equation (\ref{secondstr}),   the total variational derivative $\delta L_{TMG}$ with respect to the independent connection, coframe and auxiliary 1-forms can explicitly be expressed  as
\begin{eqnarray}
\hspace{-1em}\delta L_{TMG}
=&
-\frac{\sigma}{2} \delta\omega_{ab}\wdg D*e^{ab}
-
\delta e^c\wdg \frac{\sigma}{2}\Omega_{ab}\wdg*e^{ab}_{\fant{ab}c}
+
\delta e^a\wdg \Lambda*e_a
\nonumber\\
&+
\delta e_a\wdg D\lambda^a- \delta\omega_{ab}\wdg \frac{1}{2}\left( e^a\wdg \lambda^b-e^b\wdg \lambda^a\right)
+
\frac{1}{2\mu}
\delta \omega_{ab}\wdg \Omega^{ba}
\nonumber\\
&
+
\delta\lambda_a\wdg \Theta^a
+d
\left(
-
\frac{\sigma}{2} \delta\omega_{ab}\wdg *e^{ab}
+
\delta e_a\wdg \lambda^a
+
\frac{1}{2\mu}\delta\omega_{ab}\wdg \omega^{ba}\label{var-der-explicit-1}
\right)
\end{eqnarray}
where $D$ in the variational expression denotes the covariant exterior  derivative with respect to the independent
connection 1-form, denoted by $D(\omega)$ in the main text. The boundary term can be made to vanish by making use of the Stokes' theorem and assuming that
$\delta e^a\left|_{\partial U}\right.=0$  and $\delta \omega_{ab}\left|_{\partial U}=0\right.$.
Since $\omega_{ab}$ is assumed to be metric-compatible, $-D\eta_{ab}=\omega_{ab}+\omega_{ba}=0$ and likewise the variations satisfy $\delta\omega_{ab}+\delta\omega_{ba}=0$. Consequently the coefficient of  $\delta\omega_{ab}$ is to be antisymmetrized with respect to the indices $a$ and $b$.

The variational derivative of the $\lambda^2$-term contributes to the field equations for both the coframe and the auxiliary field  $\lambda^a$.
Explicitly, the total variational derivative of the $\lambda^2$-term  reads
\begin{eqnarray}
\delta \left(\lambda_a\wdg \lambda_b\wdg*e^{ab}\right)
&=
\delta \lambda_a\wdg \lambda_b\wdg*e^{ab}
+
\lambda_a\wdg \delta \lambda_b\wdg*e^{ab}
+
\lambda_a\wdg \lambda_b\wdg\delta*e^{ab}
\nonumber\\
&=
2\delta \lambda_a\wdg \lambda_b\wdg*e^{ab}
+
\delta e^c \wdg\epsilon_{abc}\lambda^a\wdg \lambda^b.
\end{eqnarray}

By combining the above expression for the  variational derivatives, one eventually ends up with an expression of the form
\be
\delta L_{MMG}
=
\delta e_a\wdg *E^a+\delta\omega_{ab}\wdg S^{ab}+\delta \lambda_a\wdg P^a
\ee
up to a discarded exact  3-form. The explicit expressions for the tensor-valued forms on the right-hand side read
\begin{eqnarray}
\frac{\delta L_{MMG}}{\delta e_a}
&\equiv
*E^a
=
\sigma *G^a
+
\Lambda*e^a+D\lambda^a
+
\frac{\alpha}{2}\epsilon^{a}_{\fant{a}bc}\lambda^b\wdg \lambda^c,
\\
\frac{\delta L_{MMG}}{\delta \omega_{ab}}
&\equiv
S^{ab}
=
-
\frac{\sigma}{2}D*e^{ab}
+
\frac{1}{2\mu}\Omega^{ba}
-
\frac{1}{2}(e^a\wdg\lambda^b-e^b\wdg\lambda^a),
\\
\frac{\delta L_{MMG}}{\delta \lambda_{a}}
&\equiv
P^{a}
=
\Theta^a+\alpha\epsilon^{a}_{\fant{a}bc}e^b\wdg \lambda^c,\label{constraint-appendix}
\end{eqnarray}
where  $E_a=E_{ab}e^b$ is a vector-valued 1-form, $S^{ab}=-S^{ba}$ is a antisymmetric tensor-valued 2-form, whereas $P^a$ is a vector-valued 2-form.
The constraint equation $P^a=0$ leads to a non-vanishing torsion expressed in terms of $\lambda_a$ and
it is difficult to  eliminate $\lambda_a$ from the coframe equations $*E^a=0$ by first solving the connection equations $S^{ab}=0$ for $\lambda_a$ and taking  the constraint into the account.

\end{document}